\documentclass[12pt]{article}
\newwrite\ffile\global\newcount\figno \global\figno=1

\def\writedef#1{}

\input epsf
\def\figin{\epsfcheck\figin}\def\figins{\epsfcheck\figins}
\def\epsfcheck{\ifx\epsfbox\UnDeFiNeD
\message{(NO epsf.tex, FIGURES WILL BE IGNORED)}
\gdef\figin##1{\vskip2in}\gdef\figins##1{\hskip.5in}% blank space instead
\else\message{(FIGURES WILL BE INCLUDED)}%
\gdef\figin##1{##1}\gdef\figins##1{##1}\fi}

\def\figinsert{}
\def\ifig#1#2#3{\xdef#1{fig.~\the\figno}
\writedef{#1\leftbracket fig.\noexpand~\the\figno}%
\figinsert\figin{\centerline{#3}}\medskip\centerline{\vbox{\baselineskip12pt
\advance\hsize by -1truein\center\footnotesize{  Fig.~\the\figno.} #2}}
\bigskip\endinsert\global\advance\figno by1}
\def\endinsert{}

\begin{document}
\baselineskip 18pt
\newcommand{\Tr}{\mbox{Tr\,}}
\newcommand{\beq}{\begin{equation}}
\newcommand{\eeq}{\end{equation}}
\newcommand{\bea}{\begin{eqnarray}}
\newcommand{\eea}[1]{\label{#1}\end{eqnarray}}
\renewcommand{\Re}{\mbox{Re}\,}
\renewcommand{\Im}{\mbox{Im}\,}
\begin{titlepage}

\begin{picture}(0,0)(0,0)
\put(350,0){SHEP-07-03}
\end{picture}

\begin{center}
\hfill
\vskip .4in
{\large\bf
Holographic QCD \& Perfection
}
\end{center}
\vskip .4in
\begin{center}
{\large Nick Evans}
\footnote{ I'm grateful to  J Babington, J Erdmenger, Z Guralnik,
I Kirsch, J Shock, A Tedder, and T Waterson for their contributions to the 
work reported here. }
\vskip .1in
{\em Department of Physics and Astronomy, Southampton University, \\
Southampton,
S017 1BJ, UK}

\end{center}
\vskip .4in
\begin{center} {\bf ABSTRACT} \end{center}
\begin{quotation}
\noindent A holographic description of chiral symmetry breaking
in the pattern of QCD is reviewed. D7 brane probes are used to
include quark fields in a simple non-supersymmetric deformation of
the AdS/CFT Correspondence. The axial symmetry breaking is
realized geometrically and the quark condensate and meson masses
are computable. Surprisingly, treating the model as a description
of QCD works quantitatively at the 15\% level. Models of this
AdS/QCD type typically have a strongly coupled, conformal UV
regime that is far from QCD. To systematically move closer
to QCD, we propose cutting out the large radius gravitational
description and matching operators and couplings at a finite UV
cut off in the spirit of a perfect lattice action. A simple
example is discussed. {\it Based on
a talk presented at SCGT06 in Nagoya, Japan.}
\end{quotation}
\vfill
\end{titlepage}
\eject
\noindent

\section{Introduction}

Recently the first attempts have been made to bring the
holographic techniques of the, string theory derived, AdS/CFT
Correspondence\cite{mald} to bare on QCD. The hope is that there
is some weakly coupled gravitational theory in five or more
dimensions that describes the strong coupling regime of QCD. Here
we will review a holographic description of chiral symmetry
breaking\cite{BEEGK} starting from the AdS/CFT Correspondence and
discuss to what extent it can be used as a phenomenological tool
for real QCD. 

\section{A Non-Supersymmetric Gravity Dual}

The AdS/CFT Correspondence\cite{mald} is a duality between the
conformal, large $N_c$, $N$=4 super Yang Mills theory and IIB
strings (supergravity) on 5d Anti-de-Sitter space cross a five
sphere. The field theory's global symmetries (an SO(2,4)
superconformal symmetry and an SU(4)$_R$ symmetry) match to
space-time symmetries of the $AdS$ space and the five sphere
respectively. The supergravity fields enter the field theory in
symmetry invariant ways and so appear as sources  for
field theory operators. The radial direction in $AdS$ has the
conformal symmetry properties of an energy scale and corresponds
to the renormalization group scale. Thus the radial behaviour of
the supergravity fields describes the RG flow of the field theory
sources.

Let us consider a very simple example of AdS with a scalar field,
the dilaton, switched on, due to Constable and Myers\cite{CM}
\begin{equation}
ds^2 = H^{-1/2} \left( {u^4 + b^4 \over u^4 - b^4} \right)^{\delta
/ 4} dx_4^2 + H^{1/2} \left( {u^4 + b^4 \over u^4 - b^4}
\right)^{(2-\delta) / 4} {u^4 - b^4 \over u^4}
du_6^2\end{equation} 

\begin{equation} H = \left( {u^4 + b^4 \over u^4 - b^4} \right)^\delta - 1,
\hspace{1cm} e^\Phi = \left( {u^4 + b^4 \over u^4 - b^4}
\right)^{\Delta/2}, \hspace{1cm} C_4 = H^{-1} \end{equation}

\begin{equation}
 \delta = {R^4 \over 2 b^4}, \hspace{1cm} \Delta^2 = 10 -
\delta^2 \end{equation}

The $x_4$ directions correspond to the field theory's 4d space and
$u$ is the radial direction in the 6d transverse space. At large
$u$ the space becomes $AdS_5 \times S^5$ with radius R. Here $b$
is a parameter that controls the size of the deformation from AdS
- note it enters along with $u$ and so has energy dimension one.
The SO(6) isometry of the transverse plane survives at all $u$ and
thus the R symmetry of the field theory is not broken. From these
facts we can deduce that $b^4$ corresponds in the field theory to
a vacuum expectation value for the dimension four, R-chargeless
operator $Tr F^2$.

It is worth stressing that the vacuum of the $N=4$ gauge theory
has $Tr F^2=0$ (this quantity is the D-term of a superfield and
supersymmetry would be broken were it generated) and so the above
geometry describes a non-vacuum state of the field theory. The
geometry does though describe some non-supersymmetric, strongly
coupled gauge configuration and is relatively simple - for these
virtues we will use it below. A consequence of the supersymmetry
breaking is that the dilaton (the gauge theory coupling) changes
with $u$ {\it ie} the gauge coupling runs with energy scale. It
has a pole at $u=b$ which we interpret as playing the role of the
pole in the QCD coupling at the scale $\Lambda_{QCD}$.

\section{D7 Branes and Quarks}

The $N=4$ gauge theory only has adjoint matter fields - the
original construction realized the gauge theory through open
string modes with both ends tied to a D3 brane (they transform as
$N_c, \bar{N}_c$). To generate fundamental representation quarks
one must detach one of the string's ends from the D3 - it is
useful to tie it to a D7 brane\cite{KarchKatz} as shown in Fig.
1. 

\begin{figure}[ht]
%\epsfxsize=10cm
%width of figure - will enlarge/reduce the figures
%\epsfbox{fig3.eps}
%\figurebox{2cm}{3cm}{} %to have a box alone
\centerline{\epsfxsize=2.5in\epsfbox{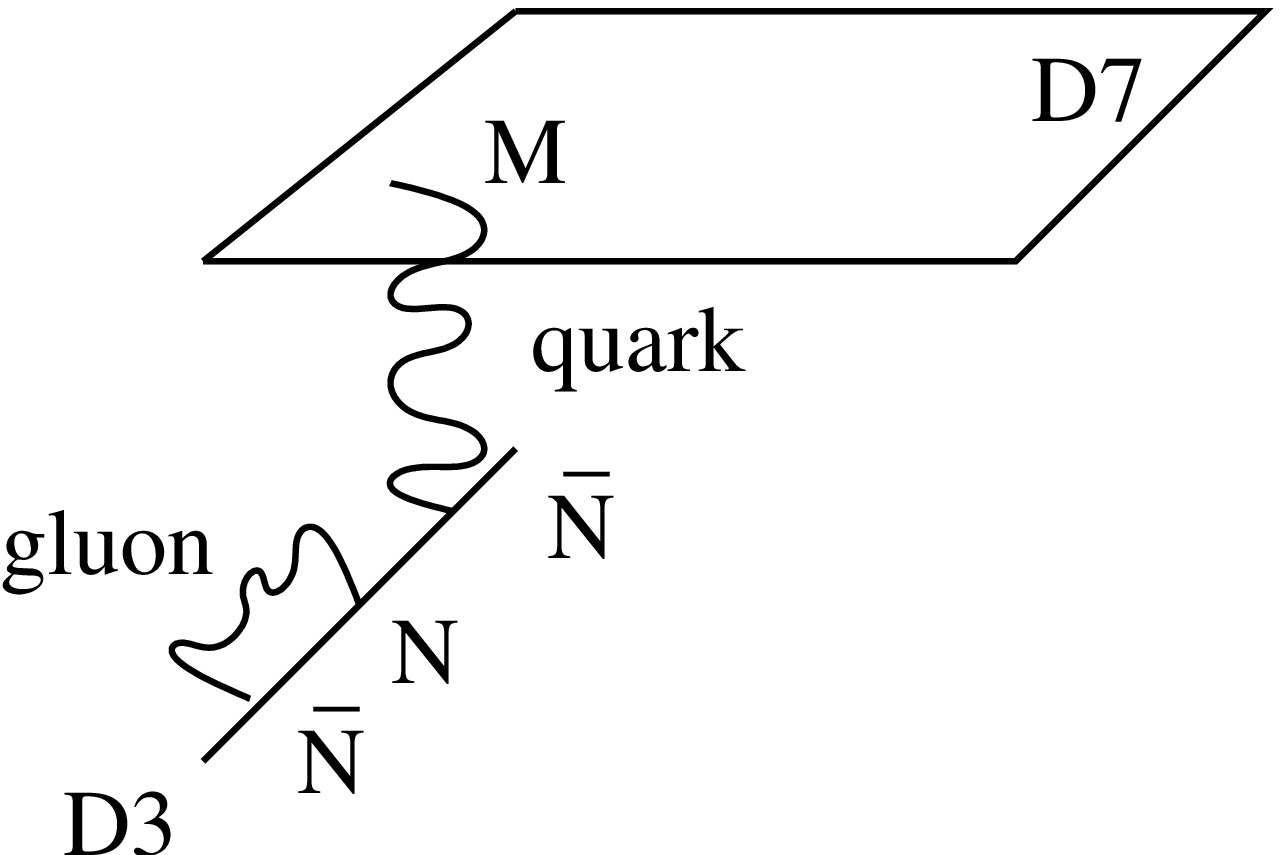} \hspace{0.5cm}
\epsfxsize=2.5in\epsfbox{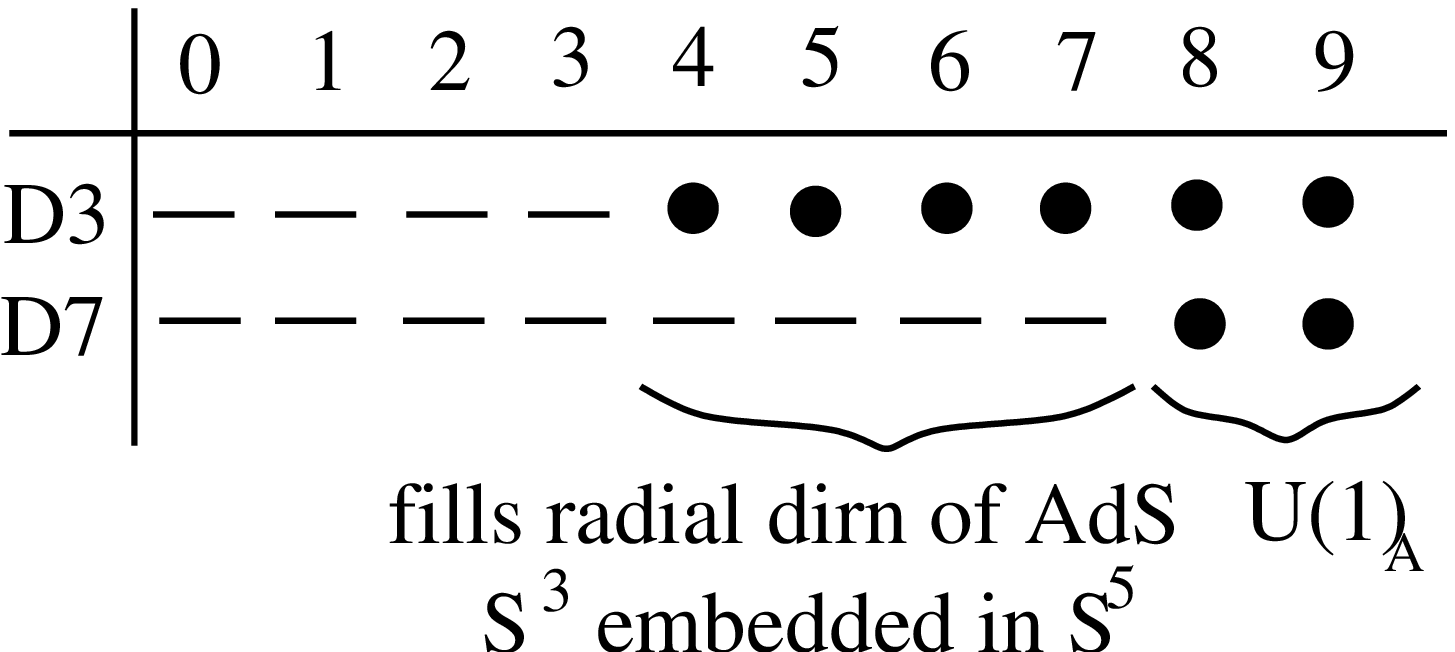}} \caption{D3/D7
configuration that introduces quarks into the AdS/CFT
Correspondence.}
\end{figure} 

The D3 and the D7 share the 0-3 directions, the D7 are in addition
extended in the 4-8 directions (we will call the radial coordinate
in this space $\rho$), and finally the D3 and D7 can be separated
in the 8-9 directions ($w_5$ and $w_6$ below). This configuration
which preserves $N=2$ supersymmetry corresponds
to the $N=4$ gauge theory with an added fundamental representation quark
hypermultiplet.

The minimum length D7-D3 string indicates (length $\times$
tension) the mass of the quark. If the D7 brane lies along the
$\rho$ axis then the quarks are massless and there is an SO(2)
symmetry in the $w_5-w_6$ plane. If the D7 lies off axis there is
a non-zero quark mass and the SO(2) symmetry is explicitly broken.
This indicates that the SO(2) symmetry is a geometric realization
of the U(1) axial symmetry of the gauge theory (in the
supersymmetric case that symmetry is part of a U(1)$_R$ symmetry).
Note that at large $N_c$ we neglect anomalies.

Using these techniques we can next include quarks into the dilaton
deformed geometry above\cite{BEEGK}. We will work in the
approximation where the D7 brane is a probe (so there is no
backreaction on the geometry) - this is the quenched limit where
the number of flavours $N_f \ll N_c$. One simply embeds the D7
brane so as to minimize its world volume via it's Dirac Born
Infeld action
\begin{equation}
S_{D7} = - T_7 \int d^8 \xi \sqrt{P[G_{ab}]}, \hspace{1cm}
P[G_{ab}] = G_{MN} {dx^M \over d \xi^a} {dx^N \over d \xi^b}
\end{equation}
where $T_7$ is the tension, $\xi$ the coordinates on the D7, $x^M$
are the spacetime coordinates and $G_{MN}$ the background metric.

In the Constable Myers geometry one finds that the D7 brane is
repelled by the singular core of the geometry and the regular
embeddings of interest are those shown in Fig. 2. 
At large $\rho$ the solutions become flat as the gauge theory
returns to AdS. The solution is of the form $w_6 = m + c/
\rho^2+..$. Here $m$ corresponds to the quark mass and $c$ to the
$\bar{q} q$ condensate - we can read off the condensate as a
function of the quark mass in this theory. A more intuitive
understanding of the embedding results from interpreting the
separation of the D7 brane from the $\rho$ axis as the effective
quark mass. As one moves in $\rho$ one is changing RG scale - at
large $\rho$ one sees a small bare quark mass but in the IR (small
$\rho$) a dynamical mass is generated.

\begin{figure}[ht]
%\epsfxsize=10cm
%width of figure - will enlarge/reduce the figures
%\epsfbox{fig3.eps}
%\figurebox{2cm}{3cm}{} %to have a box alone
\centerline{\epsfxsize=3.5in\epsfbox{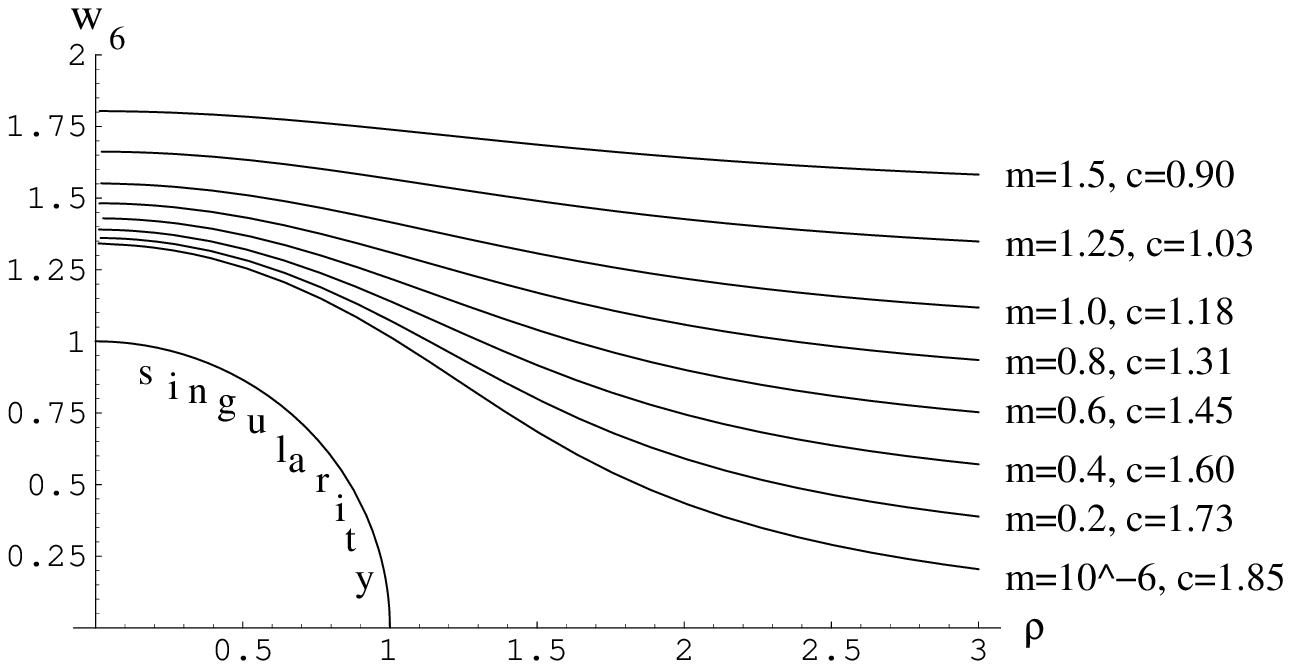} \hspace{0.5cm}
\epsfxsize=1.7in\epsfbox{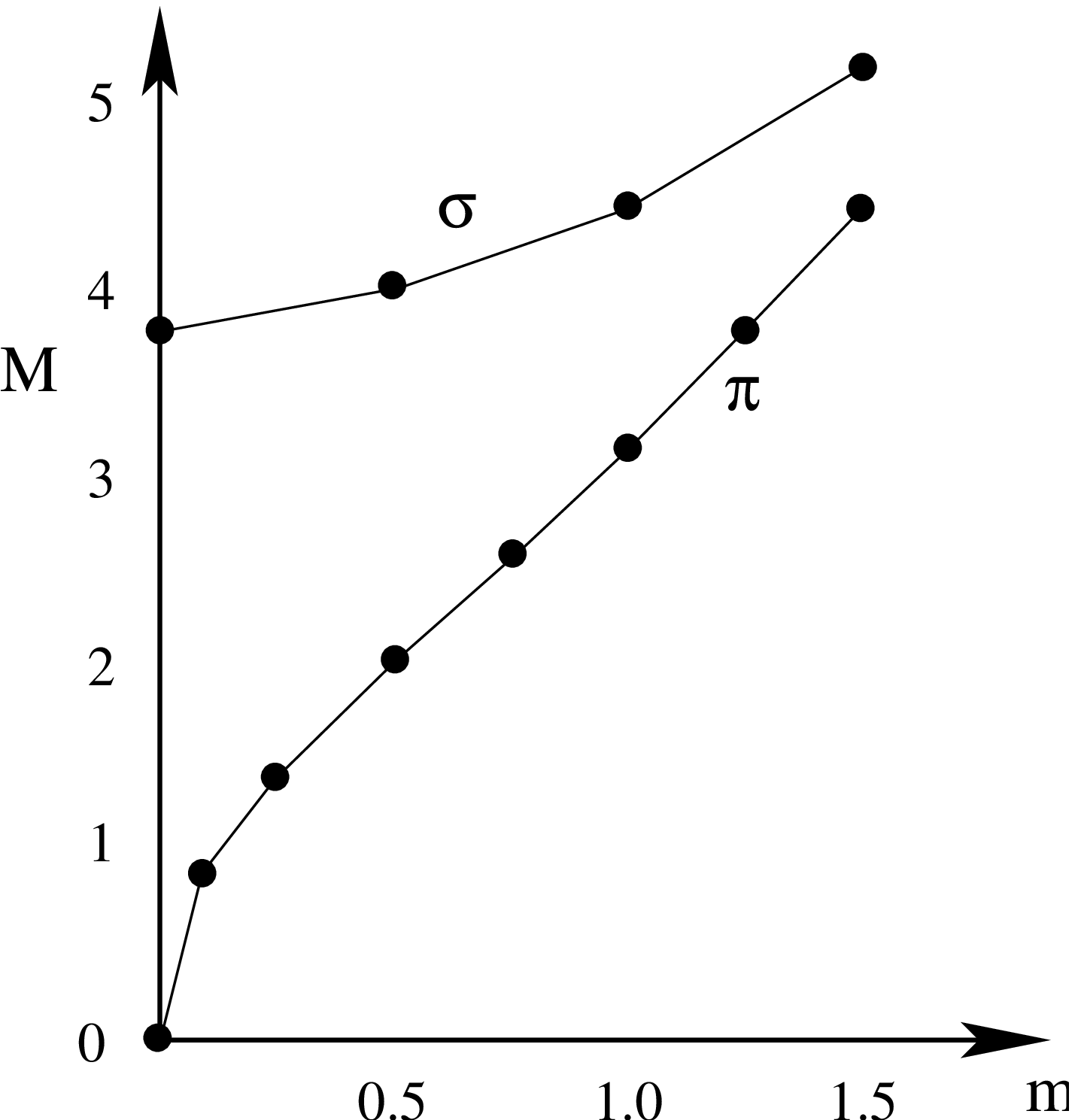}} \caption{Embedding solutions
for a D7 probe in the Constable Myers geometry and a plot of the
meson mass vs quark mass in that model.We chose $b=R$ here as
an example.}
\end{figure}

In particular we can see that the solution exhibits chiral
symmetry breaking. If we try to lie a D7 along the $\rho$ axis, so
$m=0$, it is repelled from the origin and there is a non-zero
value of the quark condensate. In fact the D7 may be deflected to
any point on a circle in the $w_5-w_6$ plane. We thus  explicitly see
the breaking of the SO(2) symmetry in that plane  and the circle
is the vacuum manifold. There should be a Goldstone boson
associated with fluctuations of the D7 along the vacuum manifold.
One can seek solutions to the equations of motion from the DBI
action for those angular fluctuations of the form
\begin{equation}
\theta(\rho,x) = f(\rho) e^{-i kx}, \hspace{1cm} k^2 = - M^2
\end{equation}
Only for particular values of $M$ is $f(\rho)$ regular and hence
the meson bound state masses are picked out. In
Fig. 2 the meson masses as a function of quark mass are shown.
There is a massless Goldstone at $m=0$ and it's mass grows as
$\sqrt{m}$ as in chiral perturbation theory. The mass of the
meson associated with radial fluctuations is also shown - it
always has a mass gap.

Note this simple model is sometimes criticized for the presence of
a singularity in the metric. It is possible a source is present at
$w=b$ that explains the singularity - one escapes addressing this
problem because the D7 never penetrates the singularity. There are
alternative D4-D6 descriptions of chiral symmetry
breaking\cite{Mateos2} that use completely smooth metrics and yet
show the same generic structure. The UV of that theory is
six dimensional though. See also other constructions in \cite{Ghoroku:2004sp}.

\section{AdS/QCD}

The holographic description of chiral symmetry breaking above
provides the pion spectrum. In addition a vector field on the D7
world volume describes the vector mesons. Solutions for these
fields with non-trivial harmonics on the $S^3$ of the D7 brane
 also exist and describe
R-charged mesons, reflecting the supersymmetric origin of the
theory. There is no significant decoupling of these R-charged
states since the theory is strongly coupled at the scale of the
supersymmetry breaking parameter, $b^4$ or $Tr F^2$.

In spite of the differences from QCD one can boldly move to a toy
model of QCD in the spirit of the work in \cite{Erlich:2005qh}
(and \cite{brodsky}). In that work a
five dimensional theory consisting of axial and vector gauge
fields and $N_f^2-1$ pions in an AdS space with a hard IR (small
$r$) cut off is studied as a model of the QCD pion, $\rho$ and $a$
mesons. We can now repeat that model but using the D7 world volume
metric from the theory above\cite{Evans:2006dj}. This has the
advantage that the conformal symmetry breaking is smoothly
included in the metric which exists down to $r=0$ rather than
through an adhoc cut off. The condensate is also a prediction of
the gauge dynamics in this model whereas it was included by hand
in the pure AdS model.

The original AdS/CFT Correspondence was for a large $N_c$ theory.
$N_c$ enters through the prediction for the relative coefficients
of the scalar and vector fields' kinetic terms 
- in the phenomenological approach
this is instead set by requiring that one reproduces the
perturbative QCD result for the vector vector correlator \cite{Erlich:2005qh}. 
Here one
is hoping that the conformal nature of the UV asymptotics of AdS
in someway mimics the conformal behaviour of weakly coupled QCD.
The remaining parameters in the model are then the conformal
symmetry breaking scale $b$ ($\Lambda_{QCD}$) and the quark mass
(position of the D7 at large $r$). Performing a global fit to
meson data one finds the results below\cite{Evans:2006dj} -
the fit is rather good (rms error 12.8\%).
\medskip

\begin{center} \begin{tabular}{ccc} \hline
& holography & expt \\
\hline
$m_\pi$ & 139.0 MeV & 139.6 MeV  \\
$m_\rho$ & 742.7MeV  & 775.8 MeV  \\
$m_a$ & 1337 MeV &  1230 MeV \\
 \hline
\end{tabular} \hspace{1cm}
\begin{tabular}{ccc} \hline
& holography & expt \\
\hline
$f_\pi$ & 83.9 MeV & 92.4 MeV \\
$f_\rho$ & 297.0 MeV & 345 MeV \\
$f_a$ & 491.4 MeV &  433 MeV\\ \hline
\end{tabular} \end{center}

\section{Perfection}

The success of the AdS/QCD approach is rather shocking - we used a
quenched, large $N_c$ gauge theory with superpartners present!
Gauge gravity dualities are also a strong weak coupling duality
and so by assuming the gravity dual is weakly coupled out to large
radius we lost QCD's asymptotic freedom. Was the success of the
fit just luck then? To answer this one must address systematic
errors - this appears hard since the theory is a model and is not
derived from QCD. Let us attempt to understand how, at least in
principle, one could make a perfect holographic description of QCD
\cite{Evans:2006ea}.

It is clear that a weakly coupled gravity description should only
exist below the scale where QCD becomes strongly coupled. We
should therefore impose a UV cut off, to represent where QCD
undergoes this transition, and work in the gravity theory only at
values of the radius below this. This is analogous to working in
lattice QCD but with a rather coarse lattice. In fact it has been
understood that one can simulate QCD on a coarse lattice and
nevertheless precisely reproduce QCD \cite{Hasenfratz:1993sp}. The
crucial point is that as one blocks from a fine lattice to a
coarse lattice one must include higher dimension operator
couplings. By analogy one should be careful to make sure all the
couplings needed to reproduce QCD are present in the gravity dual
with a UV cut off. One also needs to ensure all operators take
their appropriate vacuum value and have the correct anomalous
dimension.

In principle this is straight forward but there are an infinite
number of possible operators and couplings and
all could be large. One might worry about whether these couplings
will be sufficiently small to keep the gravity theory perturbative
- there is no guarantee but let us hope they will. In practice our
only method to fix these values is phenomenological. One might
pick on a small number of couplings and fix their values using a
fit to measured hadron data. The hope is then that those are the
significant changes needed and that the remainder of the physical
spectrum will be predicted more accurately (here the analogy is to
improving lattice actions).

As a toy example consider the AdS/QCD model in \cite{Karch:2006pv} of the
$\rho$ meson and it's excited states. The theory is just a gauge
field in AdS$_5$ with a non-zero dilaton that blows up in the IR,
$\Phi \sim r^{-2}$. In the usual approach one would fix the large
$r$ behaviour of the gauge field to enforce the $\bar{q}
\gamma^\mu q$ operator to be dimension 3 in the UV. We now though
impose that boundary condition not at infinity but at some finite
UV cut off\cite{Evans:2006ea}. In other words we ensure the
scaling dimension of the operator is three down to the scale where
QCD becomes non-perturbative. In Fig. 3 we plot the $\rho$ meson
masses for the low excitation numbers and compare to the
experimental values. Lowering the cut off improves the fit (rms error of 
2\% for the best fit). Thus changing the anomalous dimension of
the quark bilinear operators seems to be an example of an
improvement of the holographic dual. One should caution that the
importance of many other operators and couplings should  be
checked although it is far from clear how to include some of these
in the gravity dual. Hopefully though one has understood how to be
more systematic in the approach.

\begin{figure}[ht]
%\epsfxsize=10cm
%width of figure - will enlarge/reduce the figures
%\epsfbox{fig3.eps}
%\figurebox{2cm}{3cm}{} %to have a box alone
\centerline{\epsfxsize=2.5in\epsfbox{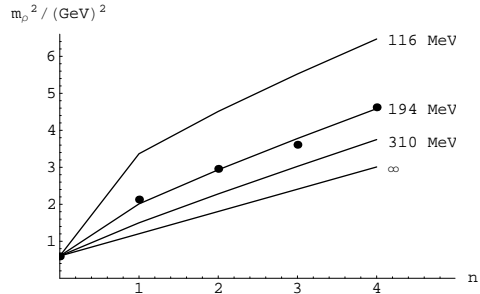}} \caption{The
$\rho$ meson and its excited states' masses with varying UV cut
off in the model in \cite{Karch:2006pv}.  The dots are the QCD data.}
\end{figure}

\end{document}